\documentclass[a4paper,11pt]{article}

\usepackage{amsmath}
\usepackage{amsfonts}
\usepackage{amssymb}
\usepackage{bbm}
\usepackage{pstricks}
\usepackage{pst-plot}
\usepackage{pst-node}
\usepackage{multido}
\usepackage[all]{xy}
\usepackage{rotating}
\usepackage[dvips]{epsfig}
\usepackage{colortbl}
\usepackage{wrapfig}
\usepackage[hang,nooneline]{subfigure}
\usepackage[footnotesize,ruled]{caption}
\usepackage{graphicx}
\usepackage{epstopdf}
\usepackage[utf8]{inputenc}
\usepackage{graphicx}
\usepackage{euscript}

\begin{document}
\title{High precision thermal modeling of complex systems with application to the flyby and Pioneer anomaly}
\author{Benny Rievers\footnote{benny.rievers@zarm.uni-bremen.de} and Claus L\"ammerzahl\footnote{claus.laemmerzahl@zarm.uni-bremen.de} \\
ZARM, University of Bremen, Am Fallturm, 28359 Bremen, Germany}

\maketitle

\begin{abstract}
Thermal modeling of complex systems faces the problems of an effective digitalization of the detailed geometry and properties of the system, calculation of the thermal flows and temperature maps, treatment of the thermal radiation including possible multiple reflections, inclusion of additional external influences, extraction of the radiation pressure from calculated surface data as well as computational effectiveness. In previous publications \cite{2009NJPh...11k3032R, Rievers2010467} the solution to these problems have been outlined and a first application to the Pioneer spacecraft have been shown. Here we like to present the application of our thermal modeling to the Rosetta flyby anomaly as well as to the Pioneer anomaly. The analysis outlines that thermal recoil pressure is not the cause of the Rosetta flyby anomaly but likely resolves the anomalous acceleration observed for Pioneer 10.
\end{abstract}

\tableofcontents

\section{Introduction}

During the last years a universal and generic tool for thermal modeling of complex systems has been developed at ZARM \cite{2009NJPh...11k3032R, Rievers2010467}. This modeling includes various components: (i) a very precise finite element (FE) model of the system, (ii) a high precision numerical computation of thermal fluxes and (iii) the determination of the resulting thermal radiation pressure (TRP). The combination of a high resolution FE modeling with complex numerical calculations enables the determination of the TRP with a very high precision. In the meantime the TRP tool has been refined and made more efficient. Here we report on two applications of this tool in the area of satellite dynamics. 

In more detail, our high precision modeling of a spacecraft and the effect of TRP includes 
\begin{itemize}\itemsep=-2pt
 \item the detailed internal and external spacecraft geometry,
 \item material models consisting of mechanical, thermal, and radiative properties (absorption, reflection, and emittance coefficients) for each component,
 \item the heat and radiation exchange between all components (internal and external),
 \item telemetry data such as temperature sensors, 
 \item ray tracing and models for re-absorption and multiple reflection of emitted heat radiation, as well as
 \item environmental conditions and the spacecraft state.
\end{itemize}
By using this procedure the TRP and other properties can be evaluated during the whole mission even in the case that the boundary loads, the material properties or the geometry of the model depend on time (e.g. through the rotation of an antenna what results in a change of the model geometry). This thermal modeling tool can be applied to complex experimental setups as well as to whole spacecraft.

The development of such a tool were motivated by the need for a precise thermal modeling for fundamental physics space missions like LISA as well as speculations that thermal effects may contribute or even may be responsible for observations known as flyby anomaly and Pioneer anomaly, see \cite{LaemmerzahlPreussDittus06}, and references therein. The anomalous acceleration of the Pioneer spacecraft has been confirmed by a number of independent data analyses \cite{Markwardt2002, Olsen2006, Levy2009, Toth2009}. There were many speculations that these anomalies may hint toward new physics or indicate an effect of the expansion of the universe on the scale of the Solar system. While the latter has been excluded \cite{HackmannLaemmerzahl08b} there were many speculations for modified gravity or modified inertial laws. However, new physics can only be seriously taken into account as possible explanation if all conventional explanations fail. Here we show that the TRP can indeed be considered as the cause for the Pioneer anomaly and also explains differences between measured and modeled orbits for the general Rosetta trajectory. However, the TRP is ruled out as the cause of the observed velocity jump during Earth flybys. 

\section{Thermal modeling and radiation pressure}

The whole procedure starts with the thermal model. This includes the build-up of a FE model of the system which can be very complex for realistic spacecraft geometries. From the environmental conditions as well as from the heat produced inside the spacecraft the temperature distribution throughout the spacecraft is calculated by means of a thermal FE analysis. Here the spacecraft geometry is modeled as a set of discrete FE which implement a system of differential equations that can be solved numerically. The results includes the temperature distribution along the surface. Here the surface has to be composed of quadrilateral 4-node surface elements with assignment of optical properties, surface temperatures and coordinates. This temperature distribution then serves as input to the TRP analysis. 


This input is processed in the TRP tool where radiation fluxes between all surface elements are treated by means of radiation shape factors which are obtained in a very high precision by numerical integration. For the assessment of multiple reflections as well as re-absorption of emitted heat a ray tracing approach is applied where each  surface element is treated as a Lambertian radiator. By applying different reflection models directly to the surfaces elements the ray directions can be computed and the radiation interaction between different surfaces can be identified. The basics of the method as well as an analytical model for the TRP have been introduced in \cite{2009NJPh...11k3032R}.

Compared with \cite{2009NJPh...11k3032R}, the present TRP computation method has been refined greatly. The computation of the radiation view factors has now been implemented by means of a Gauss quadrature which enables a very high numerical precision. Furthermore the computational performance has been increased considerably by utilizing view factor symmetries. 

%

\section{Application to Rosetta flyby}

A thorough analysis of the cause of the flyby anomaly requires a precise determination of all disturbance effects acting on the spacecraft during its gravity swing-by. For this aim our TRP computation method is used to analyse the TRP influence on the first Rosetta Earth flyby on 2005/03/05.
\subsection{The Rosetta FE model}
\begin{figure}[h!]
\centering
 \epsfig{file=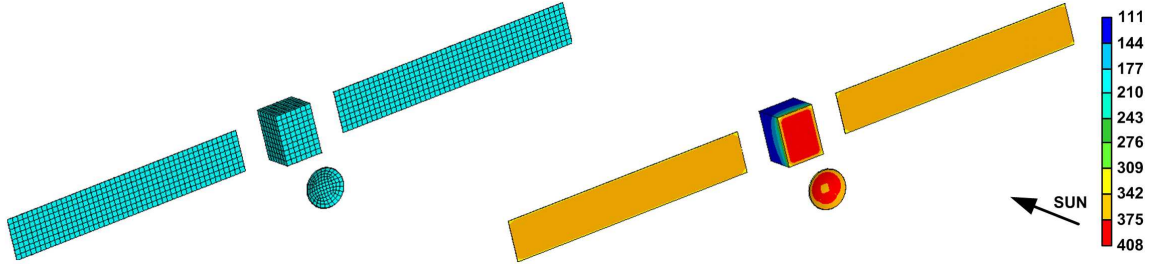, width=16.0cm, angle=0}
 \caption{Rosetta FE model (left) and resulting thermal surface temperature distribution for zero solar azimuth and elevation, as measured from bus frame (right). Temperature values are given in Kelvin. Visualisation color scale does not represent actual accuracy.} 
 \label{Rosettamodel}
\end{figure}

A sophisticated FE model of the Rosetta spacecraft consisting of over 10,000 individual FEs has been developed as displayed in Figure \ref{Rosettamodel} left. Here heat radiation between all components as well as heat conduction is implemented. The geometry of the model can be changed by a set of geometrical input parameters which account for the rotation of the solar panels or the high gain antenna (HGA).

\subsection{Flyby TRP analysis}
In order to determine the TRP acting on the spacecraft a thermal FE model of the Rosetta spacecraft has been developed. The TRP has been evaluated for chosen positions during the flyby by specifying the time dependent attitude and heliocentric distance along the trajectory. The corresponding trajectory and attitude information has been acquired from the ESA TASC website\footnote{Data has been acquired from the ESA TASC website, http://tasc.esa.int}. Then the surface temperatures are computed for each new parameter set by means of a thermal FE analysis as displayed in Figure \ref{Rosettamodel} right. For the computation of the TRP a total mass of $m_{\rm Rosetta} = 3000\;\text{kg}$ is taken into account.

\begin{figure}[h!]
\centering
 \epsfig{file=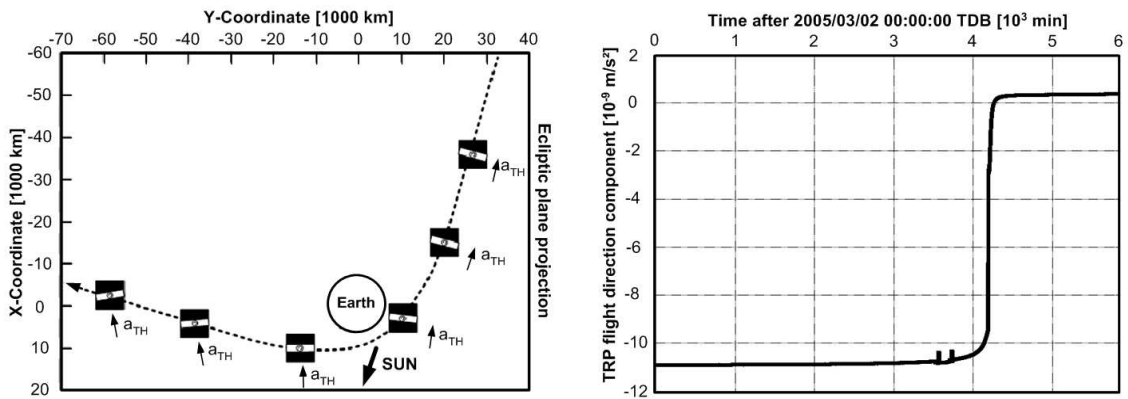, width=15.0cm, angle=0}
 \caption{The flyby of Rosetta and the TRP acting on the spacecraft during its 1st Earth flyby, with respect to flight direction.} 
 \label{Rosettaflyby}
\end{figure}

Figure \ref{Rosettaflyby} shows the characteristics of the TRP acting on the Rosetta flyby (left) and the computed magnitude of the TRP aligned with the current direction of flight (right). It can be seen that the acceleration resulting from the TRP acts against the direction of flight along the incoming branch of the flyby trajectory and changes its sign after the spacecraft passed the Earth. Two small peaks in the TRP are visible just before the spacecraft passes the planet (and the TRP changes its sign). These discontinuities indicate the performance of attitude change maneouvers which lead to a change of the bus surface exposed to the Sun and thus the TRP.

 The characteristics of the TRP direction results from the fact that the TRP is governed by the heat gradient on the Rosetta bus which is heated by the Sun. With minor directional variations the TRP is thus always aligned with the Earth--Sun direction. However, the angle between the Earth--Sun and flight direction is large for the incoming branch of the trajectory and small for the outgoing branch which leads to a higher (negative) incoming TRP and a much lower outgoing (positive) TRP. Thus, the TRP induced deceleration leads to a velocity decrease of the craft of about $2.5\cdot10^{-3}\;{\rm m/s}$ in the analyzed time interval. In contradiction to this the observed flyby anomaly has shown an increase of the spacecraft velocity. Therefore the TRP cannot be the source of the Rosetta flyby anomaly. This also complies with analytical TRP results obtained by \cite{Tempo2008}.    

\section{Application to the Pioneer anomaly}
In \cite{Andersonetal02} is has been argued that the anomalous acceleration of the Pioneer spacecraft after their last flyby at Jupiter or Saturn cannot be the result of an anisotropic thermal radiation. They considered the thermal radiation of the radioisotope thermoelectric generators (RTG) and calculated how much of this radiation can be reflected by the spacecraft. Since the solid angle of the spacecraft as seen from the RTGs is less than 2\% of $4\pi$ steradians and the RTGs mainly radiate orthogonal to the line of sight to the spacecraft, the TRP results in an acceleration less than $10^{-10}\;{\rm m/s}^2$. In the present analysis we also take into account the radiation emitted from the spaceraft itself which results in a full explanation of the anomalous acceleration. 

\subsection{The Pioneer FE model}

First TRP results with a coarse model geometry have been published in \cite{2009NJPh...11k3032R}. There we found that the TRP is a major perturbation acting on the Pioneer spacecraft and that a TRP, ranging from 120 \% to 80 \% of the Pioneer anomalous acceleration, could be explained by the coarse model approach. The model has been refined greatly since then and further telemetry data has been included into the thermal FE simulation. This results in a highly realistic temperature distribution along the surface of the spacecraft. Furthermore, the available energy of the internal compartment has been reconstructed from RTG voltage sensors, shunt voltage, cable losses and further sensor data. Numerical errors have been minimized by improving the FE mesh and a realistic internal composition of the instrument compartment with individual payload boxes, radiation exchange and conductive coupling to the ground and side panels has been implemented. A realistic multilayer insulation model (MLI) has been added which treats the different MLI sheets as additional virtual temperature degrees of freedom (DOF) at the surface nodes. The temperature distribution within the MLI is then calculated iteratively by means of the outer radiation boundary conditions and the thermal coupling to the interior spacecraft structure.

\begin{figure}[h!]
\centering
 \epsfig{file=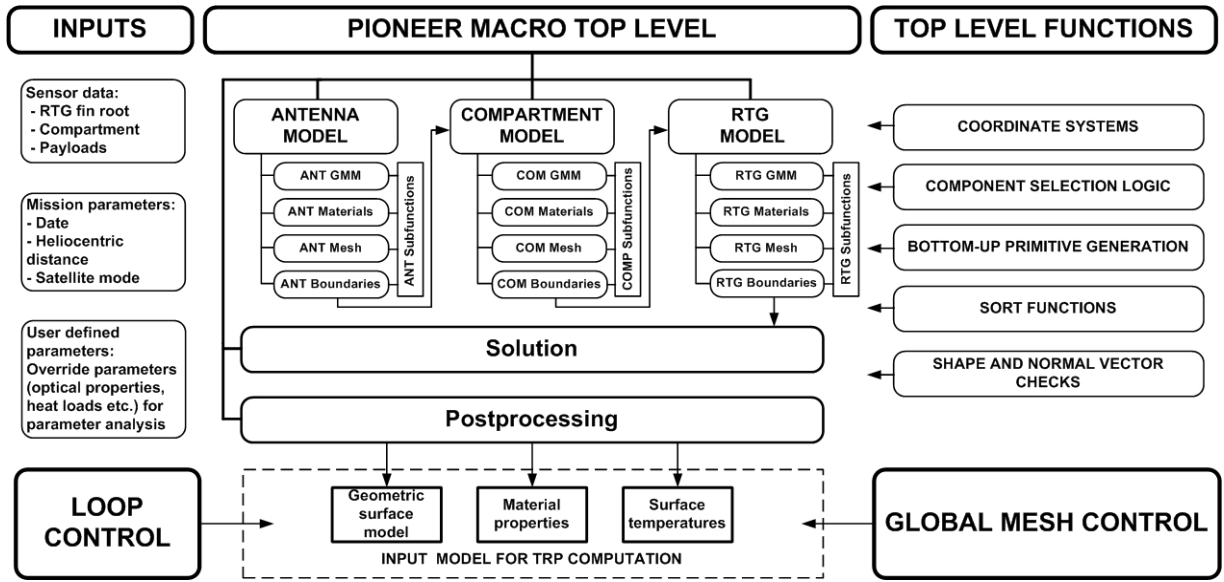, width=16.0cm, angle=0}
 \caption{Structure of the Pioneer APDL macro. By implementing a parametric approach, a high level of flexibility, mesh control and adaptability to different configurations of the thermal FE analysis is enabled.}
 \label{fig:macrostruct} 
\end{figure}

In order to enable the calculation of the distribution of the temperature along the surface of the spacecraft for the whole orbit a parametric modeling approach has to be employed. The ANSYS Parameteric Design Language (APDL) has been used to create a set of macros which process in assembly the complete thermal analysis of Pioneer 10. We defined different macro hierarchy levels which enable the control of all relevant simulation parameters by means of a single configuration macro. External loops enable the automation of the subsequent analysis with varying parameters and the export of the results into the TRP analysis tool. Auxiliary macros with standard operations such as keypoint sorting, switching to component local coordinates, or shell/solid DOF coupling have been defined to keep the complex code structure clear and comprehensible. Figure \ref{fig:macrostruct} shows the assembly of the developed Pioneer 10 FE macro and the interaction between the different macro functions and levels. 

\begin{figure}
\centering
	\epsfig{file=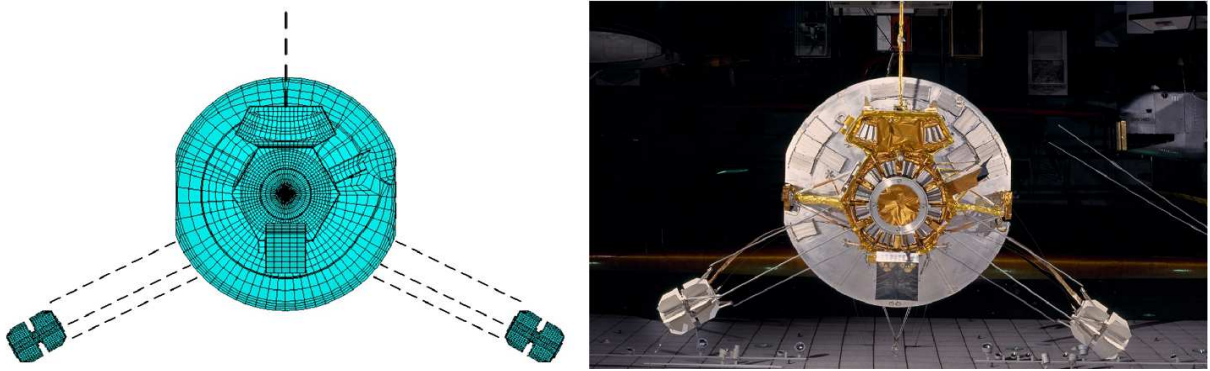, width=16cm, angle=0}
	\caption{Pioneer FE model (left) and Replica placed in the National Air and Space Museum in Washigton, DC. Note that length of the replica RTG bearings has been modified slightly with respect to the flight model.}
	\label{fig:realitycheck}
\end{figure}

The macro structure is divided into three major parts. The input part includes all macros that read data from external sources (such as heliocentric distance, temperature sensor data and energy loads) and control the parameter variation. The top level includes the actual geometry modeling, material definition, meshing and the specification of boundary conditions for each component. Here also the solution is obtained and the results are exported into the TRP tool format. The top level functions include auxiliary functions which can be used by all other macros. Finally, a global mesh control macro steers the meshing process and a loop control macro enables the configuration of subsequent simulation processing. The resulting FE mesh of the Pioneer 10 satellite and a view of the real spacecraft geometry\footnote{Picture by Smithsonian National Air and Space Musem, Washington, DC, http://www.nasm.si.edu} is displayed in Figure \ref{fig:realitycheck}.

The model implements a highly realistic geometry with external and internal payload as well as a detailed MLI model. The RTG and HGA bearing, the large magnetometer boom, guiding rods and harness (as displayed with dashed lines in Figure \ref{fig:realitycheck}) have not been included because their contribution to the resulting TRP is negligible with respect to those of the larger main structures. In total, the model consists of approximately 50,000 FEs of which 17,000 build up the surface model for the TRP computation.

\begin{figure}[h!]
\centering
 \epsfig{file=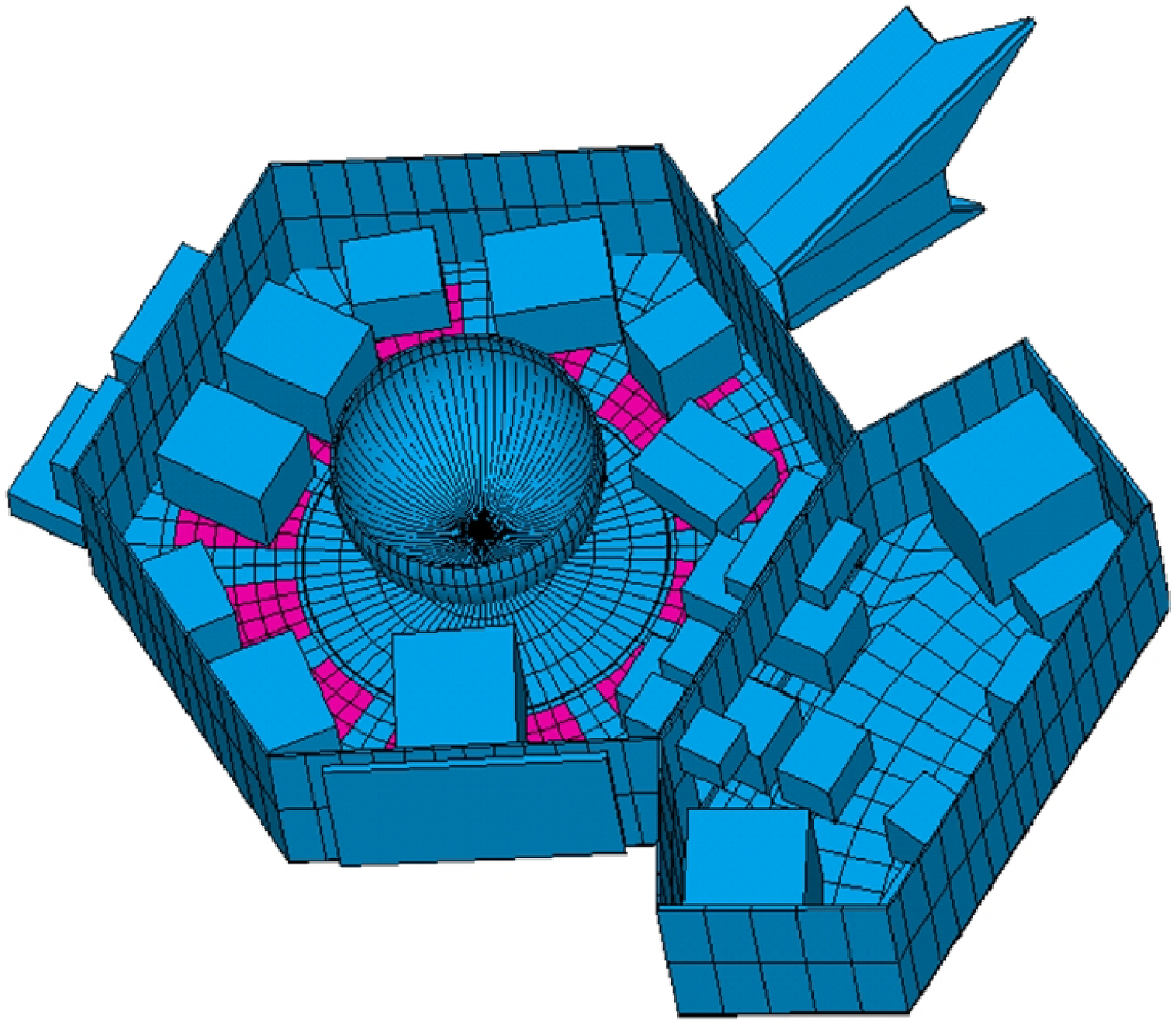, width=6.6cm, angle=0}\epsfig{file=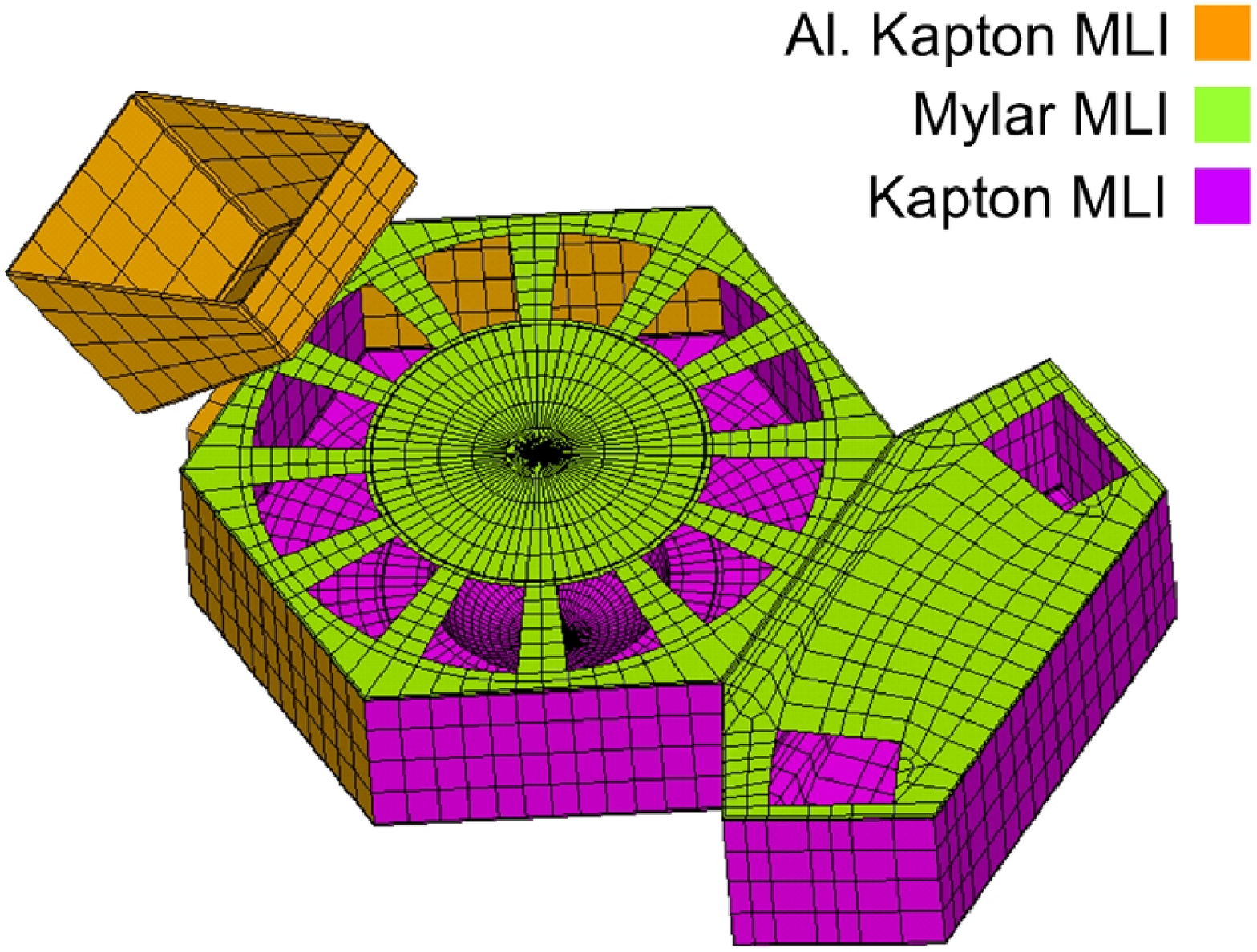, width=8.0cm, angle=0}
 \caption{Internal FE model composition (left) and MLI model.}
 \label{fig:payloadsandMLI}
\end{figure}

The MLI is implemented by means of shell FEs which effectively create additional temperature DOF on the outer surface FE nodes. Here each new DOF resembles one of the MLI sheets and their thermal coupling determines the resulting effective MLI conductivity. The inner MLI DOF is thermally connected to the underlying DOF of the solid model while the radiation boundary is specified with respect to the exterior MLI DOF. Conductive properties of the MLI are computed from the ratio of $\varepsilon^*/\varepsilon$ where $\varepsilon^*$ is the effective MLI emissivity and $\varepsilon$ is the optical emissivity of the outer MLI layer. Three different MLI materials as displayed in Figure \ref{fig:payloadsandMLI} have been implemented.

The interior payload has been modeled as individual boxes which are thermally connected to the ground and to the side panels by thermal conductors. Radiation boundary conditions have been specified on all interior free faces which allows for radiation exchange between the different components. External surfaces may radiate into space or exchange radiation with other spacecraft parts depending on their orientation and optical properties. 

\subsection{Pioneer TRP analysis}

\begin{figure}[h!]
\centering
 \epsfig{file=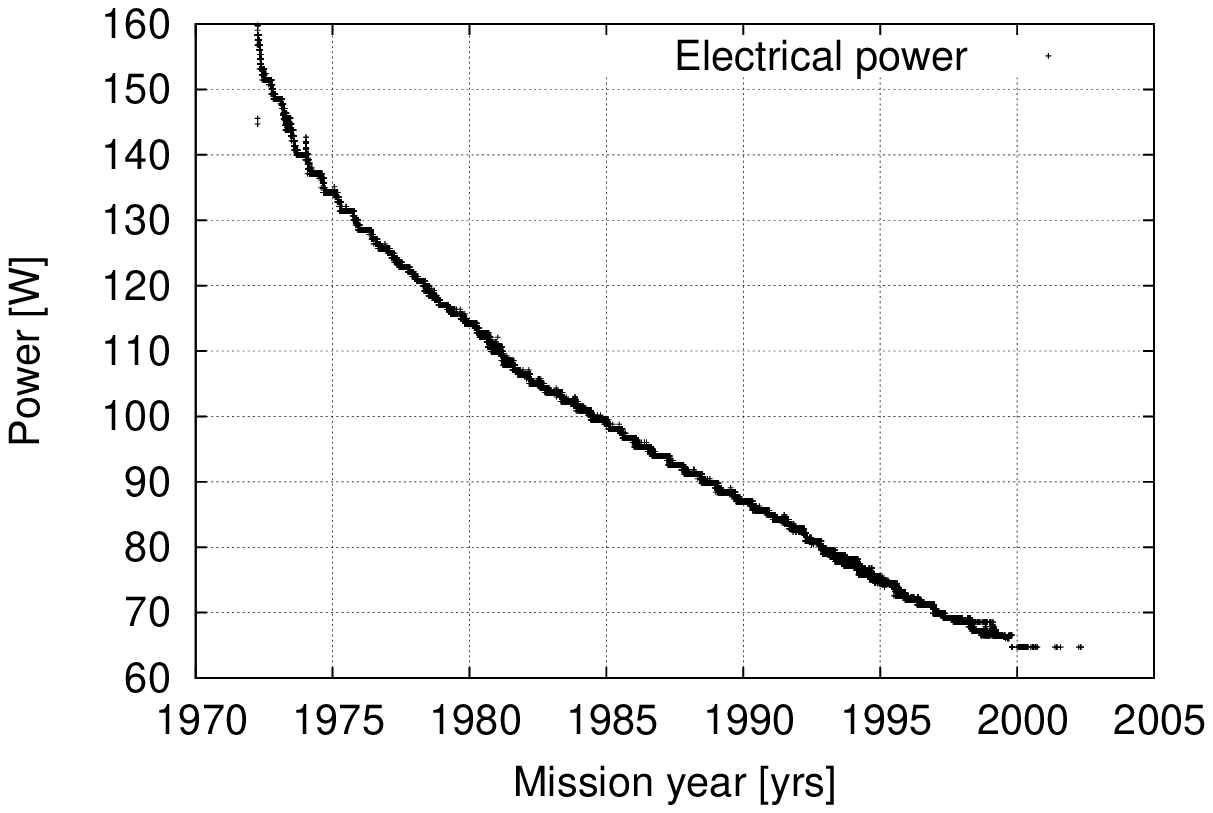, width=8cm, angle=0}\epsfig{file=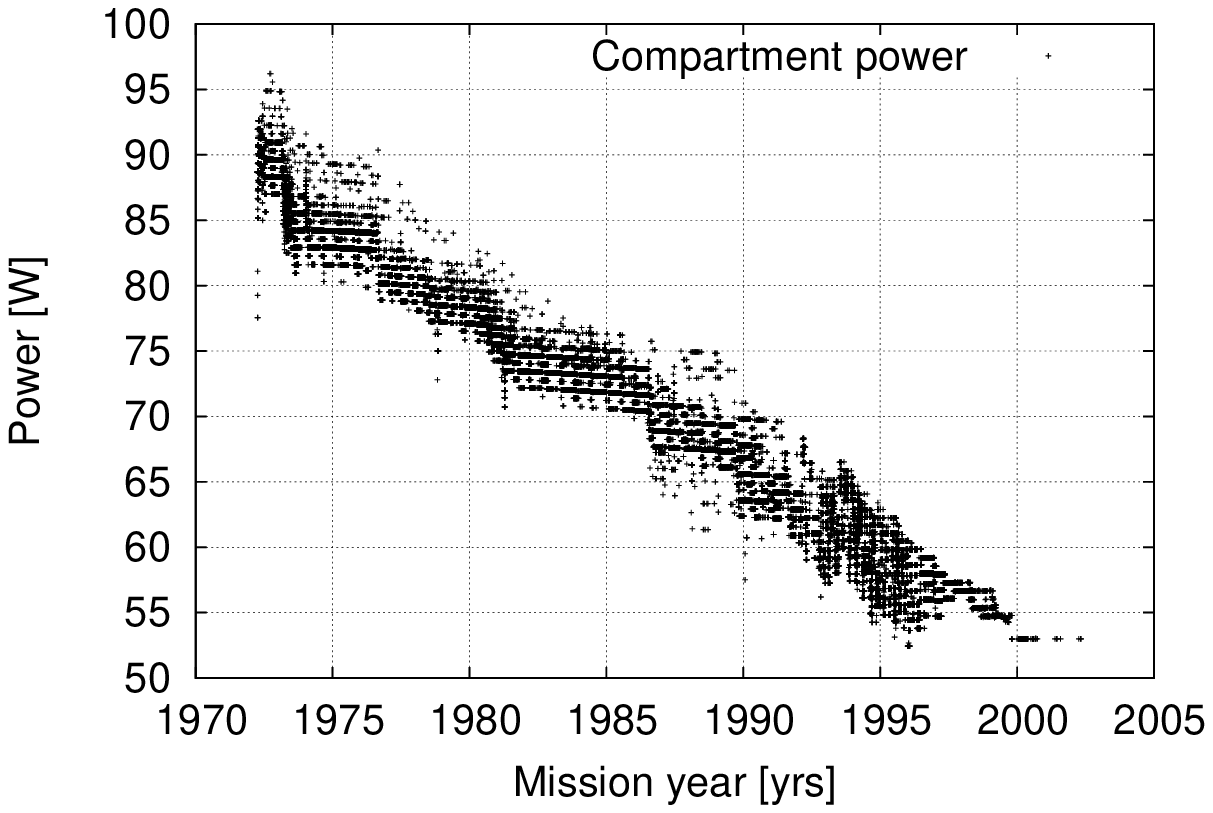, width=8.0cm, angle=0}
 \caption{Electrical power produced by RTGs (left), Electrical Power used by the compartment (right). Values are reconstructed from avalable voltage sensor readings as proposed by \cite{lrr-2010-4}.}
 \label{fig:power}
\end{figure}

We computed the TRP for the Pioneer 10 mission for the whole mission duration from 1972 to 2002 (in 2002 the last reliable data have been transmitted from Pioneer 10, the last contact was 2003 \cite{lrr-2010-4}). This corresponds to the range of approximately 1 -- 80 AU. The computations have been performed with a time resolution of 1 year. For each time the available temperature sensor data (4 fin root sensors, 6 interior compartment sensors) has been used as a constraint in the thermal FE simulation. Furthermore, heat loads on the external and internal payload, on the shunt radiator and on the high gain antenna have been specified. Internal and external heat loads were reconstructed from the available voltage sensor readings. Figure \ref{fig:power} shows the reconstructed total electrical power produced by the RTGs and the resulting power consumed by the compartment payload boxes after the extraction of power losses and power consumed by external hardware (e.g. cable losses, shunt power etc.). 

Solar illumination acting on the satellite is considered as a heat load depending on the heliocentric distance, the orientation of the surfaces, and on the occlusion by other parts of the spacecraft. For the optical surface properties estimated end--of--life (EOL) values have been considered as summarized in Table \ref {tab:opticalprops}. 

\begin{table}[h!]\centering\small
\begin{tabular}{c|c|c|c|c|c}
Component & Material & $\alpha$ & $\varepsilon$ & $\gamma_{\text s}$ & $\gamma_{\text d}$ \\
\hline
HGA front & White Paint & 0.21 & 0.84 & 0.16 & 0.01  \\
HGA rear & Bare Al 6061 & 0.17 & 0.04 & 0.96 & 0.00 \\
HGA meteroid detectors & Al 6061 & 0.36 & 0.09 & 0.81 & 0.00  \\
Louver blades (closed) & Bare Al 6061 & 0.17 & 0.04 & 0.96 & 0.00 \\
MLI & Kapton & 0.40 & 0.70 & 0.15 & 0.15 \\
 & Aluminized Kapton & 0.40 & 0.70 & 0.15 & 0.15 \\
 & Aluminized Mylar & 0.17 & 0.70 & 0.15 & 0.15\\
RTG body and fins & White Paint & 0.20 & 0.82 & 0.18 & 0.01 \\
Compartment interior & Black paint & 0.86 & 0.86 & 0.07 & 0.07 \\
\hline
\end{tabular}
\caption{Used optical properties of exterior surfaces, Values from \cite{BigBlueBook, lrr-2010-4}. In lack of actual data, the reflectivity of the MLI has been assumed to possess equal specular and diffuse components. However, the influence of MLI reflectivity on the magnitude of the TRP is very small and the actual ratio between specular and diffuse MLI reflectivity is negligible with respect to the overall TRP magnitude.}
\label{tab:opticalprops} 
\end{table}

\begin{figure}[h!]
\centering
 \epsfig{file=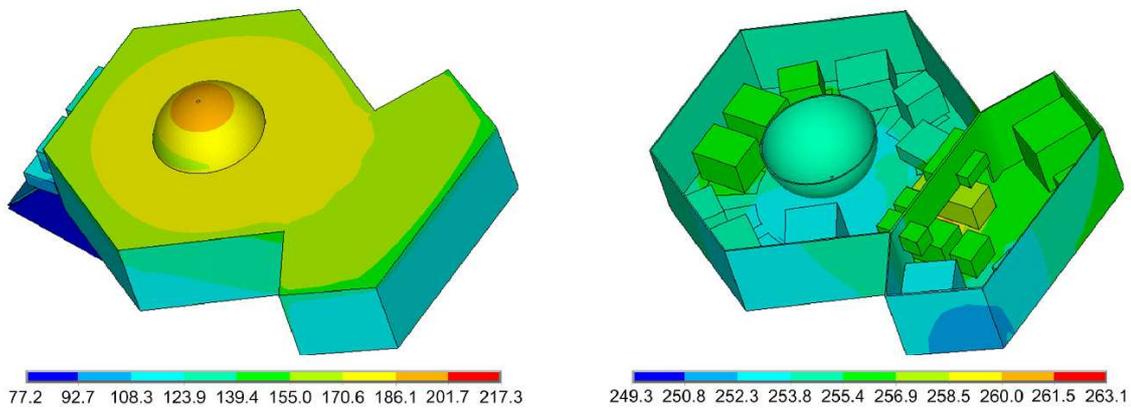, width=16cm, angle=0}
 \caption{Resulting thermal maps for external MLI sheets (left) and interior compartment temperatures (right) for conditions in 1990. Temperature values are given in Kelvin.}
 \label{fig:tempmap}
\end{figure}

Temperatures inside the craft and on the spacecraft surface have been determined for the different thermal conditions in each year of the mission. The temperatures on the outer MLI layer and inside the compartment for the year 1990 are visualized in Figure \ref{fig:tempmap}. 
It can be seen that the temperature inside the compartment is fairly constant which is a direct consequence of the insulating effect of the MLI. Compared to the inner temperatures the temperatures on the outside are much lower which also implies that heat radiated by the MLI is not a dominant effect for the total TRP since the resulting emitted fluxes are small due to its dependance on the fourth order of the temperature.

\begin{figure}[h!]
\centering
 \epsfig{file=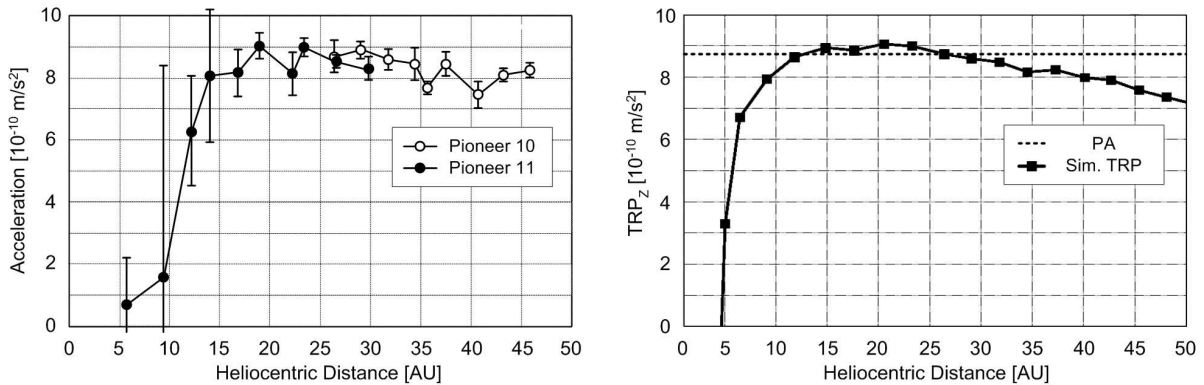, width=17.5cm, angle=0}
 \caption{Comparison between JPL determined residual acceleration \cite{lrr-2010-4} and our obtained numerical results for the TRP acting opposite flight direction of Pioneer 10. The total uncertainty is in the magnitude of 10 \% as discussed below. TRP has been evaluated for distances of 1-80 AU, however, for comparison with the JPL results only the results below 45 AU are plotted.}
 \label{fig:comparison}
\end{figure}

Figure \ref{fig:comparison} compares the residual observed acceleration determined by the JPL with our simulated results. Here the TRP component aligned with flight direction is plotted as a function of the heliocentric distance. For this a constant reference mass of 250 kg has been considered.

Our results (which have been obtained for Pioneer 10) match the observed acceleration very closely. For the early mission years the simulated TRP acts in direction of flight resulting from the close proximity to the Sun. This is caused by the highly different emissivities of the front and the rear of the HGA which effectively means that the major part of HGA heat energy is emitted by the front side, thus leading to a TRP acting against the HGA front normal.

For an increasing heliocentric distance the received Solar flux decreases quadratically and the heat produced by the RTGs and the payload becomes dominant for the TRP. Here two major contributors to the TRP can be identified. The louver blades, although completely closed for the later mission time, still emit a major part of the waste heat produced in the compartment which is mainly caused by the high insulating properties of the MLI. As the louver system is oriented into the direction of flight the emitted heat flux leads to a TRP contribution acting opposite to the flight direction. The second major TRP contribution results from the RTG fin radiation which is reflected from the rear of the HGA. Here again the resulting TRP acts against the direction of flight.

The total computed TRP thus yields an acceleration close to the anomalous Pioneer acceleration at approximately 15 -- 30 AU and decreases with a slow rate for larger distances. This behavior resembles the characteristics of both internal and RTG temperature sensors.

\subsection{Error estimation}
By means of a parameter sensitivity analysis we obtain an error estimate for our TRP result. For that the dominating parameters are the optical properties of the outer surface and the heat loads on the compartment and the RTGs. In order to quantify TRP errors inflicted by inaccuracies in the assumed optical and heat load properties we allow for small variations of the parameters as listed in Table \ref{tab:errors} and analyze their effect on the TRP. 

\begin{table}[h!]\centering\small
\begin{tabular}{c|c|c}
Parameter & Varied range & Effect on TRP \\ \hline
RTG emissivity & 0.7 - 0.9 & $\pm$ 1.5\%\\
Antenna rear reflectivity & 0.03 - 0.05 & $\pm$ 1\% \\
Louver effective emissivity & 0.03 - 0.05 & $\pm$ 1.5\%\\
MLI emissivity & 0.60 - 0.8 & $\pm$ 0.03\%\\
Compartment power & $\pm$ 10 {\rm W} & $\pm$ 1\%\\
RTG thermal power & $\pm$ 10 {\rm W} & $\pm$ 1.5\%\\
\hline
\end{tabular}
\caption{Errors determined with variation of optical properties and heat loads.}
\label{tab:errors} 
\end{table}

The ranges of the variations of the various parameters have been based on different considerations. For the optical properties different values can be found in the existing literature. Here one has to distinguish between the starting conditions (the values measured on ground) and the degraded conditions up to their EOL values. The result presented in Figure \ref{fig:comparison} has been obtained with EOL values as proposed by JPL \cite{Turyshevtalk,lrr-2010-4}. However, due to the unclear state of the degradation and the reliance on ground based measurements we take into account a possible variation of optical parameters in the range of 10\% for a conservative error estimation. 

Regarding the considered heat loads the varied range is a direct consequence of the obtained variations of the compartment power as plotted in Figure \ref{fig:power} right. Here the maximum variation from the averaged comprtment power has been considered as a worst case values. The specification of a suitable variation of total RTG power is challenging since no direct data exists. However, the coarse of the degradation of radioactive fuel can be used to determine the total power generated by the RTGs quite accurately \cite{lrr-2010-4}. The waste heat to be dissipated can then be computed by substracting the telemetred converted electrical energy from the predicted total energy. This reduces the error to the same uncertainty as discussed for the compartment power. 

As can be seen in Table \ref{tab:errors}, the computed TRP is considerably stable with respect to an input parameter variation. This directly results from the temperature measurement which have been implemented as nodal boundaries. With this input the FE system of equations is already highly constrained. If, e.g., the RTG emissivity is varied this results in a change of the temperature distribution in the unconstrained parts of the model because the total heat energy to be emitted has to remain constant. Since the emitted fluxes depend on the fourth power of the surface temperature the resulting changes in the temperature distribution are small which results in the stability of the solutions. 

Numerical errors in the FE analysis step have been minimized by mesh refinements. For doing so, additional simulations have been performed for finer mesh resolutions until a variation of 1 \% of the surface temperature between two subsequent runs was reached. Again, convergence is reached considerably fast because the model is already heavily constrained by the temperature readings. 

The numerical accuracy of the TRP computations is ensured by a large number of implemented rays per FE surface (500,000 rays per FE), together with the computation of the radiation view factor. The latter uses a numerical integration based on a Gauss quadrature with 8 points per dimension leading to a total of 128 integration points for each FE surface combination. For that the computational errors have been determined to be below 0.5 \%. 

Additional errors might be inflicted by the sensor readings themselves. However, within this study we assume them to be correct within the 6 bit resolution of the sensors One should emphasize that all temperature sensor give compatible data which minimizes the probability of erroneous measurements. Among all temperature sensors, the RTG sensor readings have the largest influence on the resulting TRP due to the higher temperatures in comparison to the compartment. For the RTGs a resulting temperature inaccuracy of approximately 1.7 K can be identified from the data resolution. The resulting error in the TRP computation decreases in the course of the mission due to its dependence on the forth power of the surface temperature. For the early mission years any TRP variation due to RTG sensor inaccuracies is hidden by the HGA TRP contribution originating from Solar illumination which acts in the opposite direction. After 10 AU, when the influence of the HGA fades out, a variation of 1.7 K of the temperature sensor results in a variation of the TRP of about 3 \%. An additional 1.5 \% TRP error can then be credited to inaccuracies in the compartment sensors.

In a worst case scenario (where all individual errors are simply added up) the total error with respect to the parameters discussed above thus evolves to $\pm$ 11.5 \%. This value has of course to be conceived as a maximum error with respect to the parameters discussed above. 

A further small inaccuracy might be introduced by the model geometry which has been implemented accurately based on the geometrical information given in \cite{BigBlueBook} and other available Pioneer documentation. However, since very detailed technical drawings for all Pioneer 10 components are not available, small deviation of model parts (e.g RTG fin dimensions) may inflict further minor inaccuracies.      

In the presented approach all surface parameter have been considered as constant during the mission. However, it is reasonable to investigate the effect of changes of the optical values during the mission. For Pioneer 10 the major fraction of surface degradation is supposed to have occured during its Jupiter encounter. Therefore it is reasonable to assume non-degraded optical properties before Jupiter and degraded EOL values after the encounter. However, due to the temperature sensor constraints the resulting characteristics of the TRP practically do not change for such a degradation profile. The resulting variations stay within the magnitude of 10 \% as discussed in Table \ref{tab:errors} and below.  


\section{Conclusion}

We have presented a high precision FE approach for the calculation of the TRP acting on a spacecraft. The generic nature of the method enables the analysis of the TRP for any spacecraft configuration and also can be applied to experimental devices and, thus, is a universal tool for the thermal modeling of complex systems. The method has been fully applied to both the Pioneer 10 and Rosetta spacecraft. In the first case the results, which are based on a detailed model of the spacecraft, flight telemetry, and the measured Pioneer 10 trajectory have shown that the observed Pioneer anomaly can be explained with an unmodeled TRP. By a variation of the parameters of the simulation we obtain that the worst case modeling error is of the order of 10 \%. With this we present the most precise assessment of TRP acting on the Pioneer 10 spacecraft to date. 

Our result concerning the Pioneer anomaly is confirmed by a recent analysis \cite{Fransiscoetal11} where a simplified model of the Pioneer spacecraft and of its heat sources has been taken into account. Furthermore, since their model does not make use of telemetry data they have to deduce the heat from the exponential decay of the power of the RTGs and to discuss effects from the shut down of systems and instruments. It turned out that the latter does not have a big influence. 

An interesting aspect is that while in \cite{Fransiscoetal11} the heat and the TRP has been calculated from the exponential decay of the power provided by the RTGs, here we used also the actually measured temperatures for the assessment of the surface temperature and the resulting TRP, both based on highly complex FE methods. While the latter result is able to determine the resulting recoil acceleration for the whole orbit, the results of \cite{Fransiscoetal11} are valid only for the later stage of the mission. For this part of the mission both results are compatible. As a consequence, the consistency of the measured anomalous acceleration based on Doppler measurements, the predictive modeling in \cite{Fransiscoetal11}, as well as our realistic model using telemetry data clearly suggests that the anomalous acceleration of the Pioneer spacecraft results from an anisotropic heat radiation. 

The analysis of the influence of the TRP on the Rosetta flyby trajectory has shown that the TRP cannot be responsible for the observed velocity jump. Nevertheless, the understanding of the direction and magnitude of the TRP during the flyby is important for any further analysis of this yet unresolved anomaly.  

\section{Acknowledgement}

We would like to thank S. Bremer, H. Dittus, E. Hackmann, M. List, S. Turyshev, and J. van der Ha for discussions and support. Part of this work has been carried out at the International Space Science Institute (ISSI), Berne, Switzerland within the scope of the Pioneer Explorer Collaboration and the International Flyby Collaboration. We thank the JPL for distributing the Pioneer Project documentation and telemetry data. We also acknowledge support by the Deutsche Forschungsgemeinschaft DFG, and the Center of Excellence QUEST.

{\footnotesize
\bibliographystyle{unsrt}
\bibliography{Bib}
}
\end{document}